\documentclass[mathleft
]{an}
\usepackage{graphicx}
\usepackage{times}
\usepackage{lscape}
\begin{document}

\Pagespan{789}{}
\Yearpublication{2006}%
\Yearsubmission{2005}%
\Month{11}%
\Volume{999}%
\Issue{88}%

\title{The DWARF project:\\
Eclipsing binaries - precise clocks to discover exoplanets}

\author{T.~Pribulla\inst{1}\fnmsep\thanks{Corresponding author:\email{pribulla@ta3.sk}\newline}
\and M.~Va\v{n}ko\inst{1} \and M.~Ammler - von Eiff\inst{2} \and M.~Andreev\inst{3,4} 
\and A. Aslant{\"u}rk\inst{5} \and N.~Awadalla\inst{6} \and D.~Balu\v{d}ansk\'y\inst{7} \and A.~Bonanno\inst{8} 
\and H.~Bo\v{z}i\'c\inst{9} \and G.~Catanzaro\inst{8} \and L.~\c{C}elik\inst{10,11}
\and P.E.~Christopoulou\inst{12} \and E.~Covino\inst{13} \and F.~Cusano\inst{14} \and D.~Dimitrov\inst{15} 
\and P.~Dubovsk\'y\inst{16} \and 
\and E.M.~Esmer\inst{10,11} \and A.~Frasca\inst{8} \and \v{L}.~Hamb\'alek\inst{1} 
\and M.~Hanna\inst{6} \and A.~Hanslmeier\inst{17} \and B.~Kalomeni\inst{18} \and D. P.~Kjurkchieva\inst{19} 
\and V.~Krushevska\inst{20} \and I.~Kudzej\inst{16} \and E.~Kundra\inst{1} \and Yu.~Kuznyetsova\inst{20} 
\and J.W.~Lee\inst{21} \and  M.~Leitzinger\inst{17} \and G.~Maciejewski\inst{22} 
\and D.~Moldovan\inst{23} \and M.H.M.~Morais\inst{24} \and M.~Mugrauer\inst{25}
\and R.~Neuh\"auser\inst{25} \and A.~Niedzielski\inst{22} 
\and P.~Odert\inst{17} \and J.~Ohlert\inst{26,27} \and \.{I}.~{\"O}zavc{\i}\inst{10,11}
\and A. Papageorgiou\inst{12} \and \v{S}.~Parimucha\inst{28} \and S.~Poddan\'y\inst{29,30} 
\and A.~Pop\inst{23} \and M.~Raetz\inst{31} \and S.~Raetz\inst{25} \and Ya.~Romanyuk\inst{4} 
\and D.~Ru\v{z}djak\inst{9} \and J.~Schulz\inst{32} \and H.V.~\c{S}enavc{\i}\inst{10,11} \and T.~Szalai\inst{33} 
\and P.~Sz\'ekely\inst{34} \and D.~Sudar\inst{9} \and C.T.~Tezcan\inst{10,11} \and M.E.~T{\"o}r{\"u}n\inst{10,11} 
\and V.~Turcu\inst{23} \and O.~Vince\inst{35} \and M.~Zejda\inst{35}
}
\titlerunning{Project DWARF}
\authorrunning{T.~Pribulla et al.}
\institute{
Astronomical Institute, Slovak Academy of Sciences, 059~60 Tatransk\'a Lomnica,
Slovakia 
\and
Th\"{u}ringer Landessternwarte Tautenburg, Sternwarte 5, 07778 Tautenburg, Germany 
\and 
Terskol Branch of INASAN RAS,  81 Elbrus ave., ap. 33, Tyrnyauz,
Kabardino-Balkaria Republic, 361623 Russian Federation 
\and
International Centre for Astronomical, Medical and Ecological
Research,  National Academy of Sciences of Ukraine, 27 Akademika
Zabolotnoho St, 03680 Kyiv, Ukraine 
\and
University of Ondokuz May{\i}s, Faculty of Science, Dept. of Physics, 55139
Kurupelit-Samsun Turkey 
\and
National Research Institute of Astronomy, and Geophysics, Helwan, Cairo, Egypt 
\and
Roztoky observatory, Slovakia 
\and
INAF, Osservatorio Astrofisico di Catania, via S. Sofia, 78, 95123 Catania, Italy 
\and
Hvar Observatory, Faculty of Geodesy, University of Zagreb, 10000 Zagreb, Croatia 
\and
University of Ankara, Faculty of Science, Dept. of Astronomy and
Space Sciences, 06100 Tandogan-Ankara, Turkey 
\and
Institute of Theoretical and Applied Physics (ITAP)
48740 Turun\c{c}, Mu\u{g}la, Turkey 
\and
Astronomical Laboratory, Dept. of Physics, University of Patras, 26500 Rio-Patras, Greece 
\and
INAF, Osservatorio Astronomico di Capodimonte, via Moiariello 16, 80131 Napoli, Italy 
\and
Osservatorio Astronomico di Bologna, Via Ranzani 1, 40127 Bologna, Italy 
\and
Institute of Astronomy and National Astronomical Observatory, Bulgarian 
Academy of Sciences, 72 Tsarigradsko Shosse Blvd., 1784 Sofia, Bulgaria
\and
Vihorlat Observatory, Mierova 4, 066 01 Humenn\'e, Slovakia 
\and
Institut f\"ur Physik, IGAM, Karl-Franzens Universit\"at Graz, Universit\"atsplatz 5, 8010 Graz, Austria 
\and
Dept. of Astronomy and Space Sciences, University of Ege, 35100 Bornova, Izmir, Turkey 
\and
Dept. of Physics, Shumen University, 9700 Shumen, Bulgaria 
\and
Main Astronomical Observatory of National Academy of Sciences of Ukraine,
27 Akademika Zabolotnoho St, 03680 Kyiv, Ukraine 
\and
Korea Astronomy and Space Science Institute (KASI), 776 Daedeokdae-ro, Yuseong-gu, Daejeon 305-348, Korea 
\and
Toru\'n Centre for Astronomy, N. Copernicus University, Gagarina 11, 87100, Toru\'n, Poland 
\and
Astronomical Institute of the Romanian Academy, Astronomical Observatory 
Cluj-Napoca, Str. Ciresilor 19, 400487 Cluj-Napoca, Romania
\and
Department of Physics \& I3N, University of Aveiro, Campus 
Universit\'ario de Santiago, 3810-193 Aveiro, Portugal
\and
Astrophysikalisches Institut und Universit\"ats-Sternwarte Jena, 
Schillerg\"a\ss chen 2, 07745 Jena, Germany
\and
University of Applied Sciences, Wilhelm-Leuschner-Stra\ss e 13, 61169 
Friedberg, Germany
\and
Michael Adrian Observatory, Astronomie Stiftung Trebur, Fichtenstrasse 
7, 65468 Trebur, Germany
\and
Institute of Physics, Faculty of Natural Sciences, \v{S}af\'arik University, Ko\v{s}ice, Slovakia 
\and
Astronomical Institute, Faculty of Mathematics and Physics, Charles University Prague, 
V Hole\v{s}ovi\v{c}k\'ach 2, 180 00 Prague 8, Czech Republic
\and 
\v{S}tef\'anik Observatory, Strahovsk\'a 205, Prague, Czech Republic 
\and
Private observatory, Stiller Berg 6, 98587 Herges-Hallenberg, Germany 
\and
Volkssternwarte Kirchheim Arnstaedter Stra\ss e 49, 99334 Kirchheim, Germany 
\and
Dept. of Optics and Quantum Electronics, University of Szeged, D\'om t\'er 9, 6720 Szeged, Hungary 
\and
Dept. of Experimental Physics, University of Szeged, Hungary 
\and
Astronomical Observatory, Volgina 7, 11060 Belgrade, Serbia 
\and
Dept. of Theoretical Physics and Astrophysics, Masaryk University, 
Kotl\'a\v{r}sk\'a 2, 611 37, Brno, Czech Republic
}

\newpage

\received{30 Mar 2012}
\accepted{11 Nov 2012}
\publonline{later}


\abstract{We present a new observational campaign, DWARF, aimed at detection of circumbinary extrasolar planets using
the timing of the minima of low-mass eclipsing binaries. The observations will be performed within an extensive 
network of relatively small to medium-size telescopes with apertures of $\sim$~20~--~200~cm.  The starting sample of the objects 
to be monitored contains (i) low-mass eclipsing binaries with M and K components, (ii) short-period binaries 
with sdB or sdO component, and (iii) post-common-envelope systems containing a WD, which enable to determine 
minima with high precision. Since the amplitude of the timing signal increases with the orbital period of an
invisible third component, the timescale of project is long, at least 5-10 years.
The paper gives simple formulas to estimate suitability of individual eclipsing binaries for the circumbinary
planet detection. Intrinsic variability of the binaries (photospheric spots, flares, pulsation etc.) limiting
the accuracy of the minima timing is also discussed. The manuscript also describes the best observing strategy 
and methods to detect cyclic timing variability in the minima times indicating presence of circumbinary planets.  
First test observation of the selected targets are presented.
}

\maketitle

\section{Introduction}

With the continuing discovery of extrasolar planets and an expectation that the majority of solar-type stars
reside in binary or multiple systems (Duquennoy \& Mayor 1991), planetary formation in binary systems has become
an important issue (Lee et al. 2009). In general, we can consider planetary companions to binary stars in two
types of hierarchical planet-binary configurations: first "S-type" planets which orbit just one of the stars,
with the binary period being much longer than that of the planet; second, "P-type" or circumbinary planets,
where the planet simultaneously orbits both stars, and the planetary orbital period is much longer than that
of the binary (Muterspaugh et al. 2007). Simulations show either of above possibilities has a large range of
stable configurations (see e.g., Broucke 2001; Pilat-Lohinger \& Dvorak 2002; Pilat-Lohinger et al. 2003;
Benest 2003). Recent theoretical studies (e.g., Moriwaki \& Nakagawa 2004; Quintana \& Lissauer 2006;
Pierens \& Nelson 2008) have predicted that P-type planets can form and survive over long timescales. Characterization of such
planets is potentially of great interest because they can lead to a better understanding of the formation and 
evolution of planetary systems around close binary stars, which can be rather different from the case of single stars.
(Lee et al. 2009). Hereafter we will consider the "P-type" or circumbinary planets only.

The detection of circumbinary planets is far from being easy. Three principal techniques are: (i) precise
radial-velocity (hereafter RV) measurements to detect the wobble of the binary mass center (Konacki et al. 2009),
(ii) photometric detection of transits of the planet(s) across the disks of the components of the inner binary (Doyle et al. 2011),
and (iii) timing of the inner binary eclipses (Lee et al. 2009).

The classical RV technique is complicated by the fact that the large RV changes of the binary components mask
the small wobble resulting from the circumbinary planet(s). In the case of close binaries, the situation is
exacerbated by the tidal spin-up of the components, where the projected rotational velocity ($v \sin i$) often
exceeds 100 km~s$^{-1}$ (see the DDO close-binary project, Pribulla et al., 2009 and references therein). This
makes the RV precision insufficient to detect any systemic velocity changes. In fact, there are hardly any binaries
where the systemic-velocity changes revealed a third component (unseen in spectra). The second technique - to detect
circumbinary planets searching for transits of a substellar companion across a close binary - requires
very long photometric runs with excellent accuracy. Assuming that the orbital planes of the underlying binary
and the outer orbit of the substellar body are close to being coplanar, the method should be advantageous for edge-on
eclipsing binaries (EBs). Three such systems were found in the Kepler satellite photometry: Kepler-16b
(Doyle et al., 2011), Kepler-34b and Kepler-35b (Welsh et al. 2012). The latter two systems were actually
identified by eclipse timing. Even if the substellar component is not 
transiting the inner binary, it causes timing variations of eclipses of the binary system due to the finite 
velocity of light (light-time effect, hereafter LITE). The eclipses act as an accurate clock for detecting other 
objects revolving around the inner binary and to determine their orbital parameters from the 
$observed-calculated$ times of minima (O-C) in way similar to the solution of RV curves. The timing technique 
proved to be the most fruitful in detecting circumbinary planets (see Section~\ref{timing_searches}).

The principal goal of the project DWARF is to detect circumbinary planets and/or substellar companions through the timing
analysis of selected close eclipsing binaries.

In this paper, we first summarize previous and/or ongoing searches for timing variability (Section~\ref{timing_searches})
and then describe the target selection criteria for the DWARF project (Section~\ref{sample}). In Section~\ref{reduction}
we present the CCD data reduction technique and determination of minima times. The following section describes the 
modeling of the (O-C) residuals for the observed targets. The telescope network put together for continuous photometric
monitoring and follow-up observations is described in Section~\ref{network}.

\section{Previous and ongoing searches of circumbinary exoplanets by timing}
\label{timing_searches}

The timing technique requires to precisely measure the exact instant of some well-defined and repeating feature of the
binary-star light curve (LC). It can be pulse arrival time in pulsar binaries or center of minimum in classical
eclipsing binaries\footnote{In principle one could use maxima of pulsations in binary stars with pulsating
component(s).}. The former technique proved to be very sensitive and led to detection of the first extrasolar
planetary system (PSR1257+12, Wolszczan \& Frail, 1992). The latter method does not require high-end
instrumentation. It will be utilized in the present observing project.

Timing of the eclipses in binary stars served as a tool to detect unseen components for several decades (see
Pribulla \& Rucinski, 2006; Beuermann et al. 2011). Unlike the RV variations, the
amplitude of the observed LITE increases with orbital period (see equation~\ref{lite_amp}). The technique
is, therefore, sensitive to substellar bodies on long-period orbits. With timing accuracies of about $\pm$ 10s
for selected EBs showing sharp eclipses, it should be possible to detect circumbinary planets
of $\sim$~10$M_J$ in long-period orbits of 10 -- 20 yr (Ribas 2006).

In the past two decades, the eclipse timing has been used to infer the existence of multiple low-mass planetary
objects to a couple of binaries. Early observations focused at the M-dwarf EB CM~Dra (see
Table~\ref{sampletab}) which was a very promising target because of deep and narrow eclipses and low total mass.
The existence of a planetary system around CM~Dra is still dubious (see Deeg et al. 2008; Ofir 2008).
A two-planet system orbiting HW Vir, a short-period EB composed of an sdB and an M dwarf, was found by Lee et al. (2009).

Post-common-envelope systems containing a WD and a main-sequence late-type dwarf enable very precise timing
because of short ingress and egress durations. The brightest such system, V471~Tau, shows timing variations
indicating the presence of a brown-dwarf companion (see Kundra \& Hric, 2011). Recently, circumbinary planets were
announced around a couple of post-common-envelope systems: two planets in possible mean-motion resonances around
the deeply eclipsing binary NN~Ser (Beuermann et al. 2010), two giant planets orbiting UZ~For (Potter et al. 2011) and
a single planet around DP~Leo (Qian et al. 2010), HU~Aqr (Qian et al. 2011), and RR~Cae (Qian et al. 2012).
In the case of HU~Aqr Gozdziewski et al. (2012) recently showed that the timing data are more consistent with
a single planet orbiting the parent star.
The exoplanet encyclopedia\footnote{\tt http://www.exoplanet.eu/catalog.php} lists (as of April 20, 2012) 11 planetary
systems (16 planets/4 multiple planet systems) detected by timing. This, however, includes a planet around
the {\it single} pulsating variable star V391~Peg found by the timing technique (Silvotti et al., 2007).

All previous monitoring of EBs focused to find exoplanets by the timing of minima was mostly
carried out by individual research groups and telescopes (e.g., Lee et al. 2009; Beuermann et al. 2011; Qian et al.
2011, 2012). An observing campaign to monitor NN~Ser over the first half of 2010 was organized by the G\"ottingen,
McDonald, and Warwick research groups (see Beuermann et al. 2010). The only larger initiative, we are aware of, is
the Polish project SOLARIS to detect circumbinary planets in the Southern sky (Konacki et al. 2011). The group is
establishing a global network of four 50cm robotic telescopes (Australia, Africa and South America) to collect
high-precision, high-cadence LCs of selected Southern binaries. The goal for timing precision should
be one second per eclipse\footnote{\tt http://www.projektsolaris.pl/photo/1538.html}.

Because the timing technique is sensitive to extrasolar planets on long-period orbits, the archival data play
an important role. A very useful source of the minima timings is the Krakow database prepared and frequently updated
by Prof. Kreiner\footnote{\tt http://www.as.up.krakow.pl/ephem/}. The major problem when using published timings
is the inhomogeneity of the data mostly caused by different approaches to determine the minima. The original LCs
are hard to come by. The situation is exacerbated by many mistakes such as heliocentric correction missing,
time shifted by one hour or typos. The minima uncertainties are often missing or underestimated.

The presented project has several advantages compared to other studies. Namely: (i) it is complementary 
to the competing project SOLARIS covering the Southern hemisphere, (ii) it uses only existing facilities, 
(iii) unlike other projects it would cover relatively extensive list of targets increasing chance of new detection(s),
(iv) it is a unique collaboration of many observatories including well equipped amateurs.

\section{Target selection}
\label{sample}

To get the highest possible accuracy and precision of the eclipse timings necessary to detect exoplanets, we
selected objects with sharp and deep minima. The following three groups of objects were included:
(i) systems with K or/and M dwarf components (ii) systems with hot subdwarf (sdB or sdO) and K or M dwarf components
(iii) post-common-envelope systems with a white dwarf (WD) component. Contact binaries will not be observed
within this campaign because of the strong interaction of the components in the common envelope
that introduces noise in their light curves. Moreover, their minima are normally broader and the 
ingress/egress phases less steep than in the previous class of EBs.

In addition to well-studied targets, we will perform follow-up observations of recently discovered detached EBs
based on the NSVS data (Hoffman et al., 2008), HAT network data (Hartman et al., 2011) and ASAS data (see Pojmanski,
2002; Pojmanski, 2003, Pojmanski \& Maciejewski, 2004ab; Pojmanski et al., 2005).

Because all observatories participating in the campaign are north of the 30th parallel, we limit the objects
to systems North of DEC = $-10\degr$. To collect as many minima as possible, and to fully cover a minimum within
one night from mid-latitudes we excluded objects with orbital periods longer than 5 days. The brightness range
of our preliminary sample is $R$ = 10-17mag, which fits the possibilities of small telescopes with apertures
of 20 -- 200cm equipped with a low-end CCD camera and at least the $VRI$ filter set. Such instrumentation
allows us to extend the observing network to well-equipped amateur astronomers. The preliminary target list
is given in Table~\ref{sampletab}.

\begin{table*}
\scriptsize
\tabcolsep 2.5pt
\caption{Target list \label{sampletab}}
\begin{center}
\begin{tabular}{lcccclcclllllrrrrc}
\hline
\hline
Target            &$\alpha_{2000}$ & $\delta_{2000}$ & $M_1$     & $M_2$      & Sp.type  & $A_{OCE}$   & $\Delta T$ &  HJD$_{I}$   & Period  & $d_I$ & $d_{II}$ & $D_I$  & $V$ & $R$ & $\Delta t$ & $N_\sigma$ & Ref.  \\
                  &                &                 &[M$_\odot$]& [M$_\odot$]&          & [mag] & [s]        & 2\,400\,000+ & [days]  & [mag] &  [mag]   & [days] &     &     & [s]        &            &       \\
\hline
DV Psc            & 00 13 09.2 & +05 35 43 &   0.49&   0.51& K5V+M1V   &  0.04 &   4.5&  52500.1150&   0.30853740& 0.32&   0.15&   0.062 & 10.6& 10.0&   1.1&   4.1& (1) \\
PTFEB11.441       & 00 45 46.0 & +41 50 30 &   0.51&   0.35& M3.5+WD   &       &   5.0&  55438.3165&   0.35871000& {\it 0.20} & {\it 0.00}&         &     & 16.3&      &      & (2) \\
NSVS 06507557     & 01 58 23.9 & +25 21 20 &   0.66&   0.28& K9+M3     &       &   4.7&  54746.3801&   0.51508836& 0.70&   0.23&   0.062 & 13.4& 12.6&   1.8&   2.6& (3) \\
BX Tri            & 02 20 50.8 & +33 20 48 &   0.51&   0.26& M1V+M4V   &  0.03 &   5.4&  51352.0616&   0.19263590& 0.33&   0.27&   0.072 & 13.4& 12.5&   4.2&   1.3& (4) \\
V449 Per          & 02 57 33.5 & +35 14 01 &       &       &           &       &      &  52500.9069&   0.94620690&     &       &         &     & 12.5&      &      & (5) \\
GJ 3236           & 03 37 14.1 & +69 10 50 &   0.38&   0.28& M4V       &  0.02 &   5.9&  54734.9959&   0.77126000& 0.21&   0.19&   0.039 & 14.0& 13.5&   6.3&   0.9& (6) \\
V912 Per          & 03 44 32.2 & +39 59 35 &       &       &           &       &      &  53287.8523&   0.57759120& 0.22&   0.20&         & 13.1&     &      &      & (9) \\
NLTT11748         & 03 45 16.8 & +17 48 09 &   0.28&   0.27& WD+WD     &       &   6.6&  55619.4264&   0.11601549& {\it 0.55}&  {\it 0.11}&         & 16.7& 16.3&      &      & (7) \\
V471 Tau          & 03 50 25.0 & +17 14 47 &       &       & K2V+DA    &       &      &  52500.3434&   0.52118357& {\it 0.03}&  {\it 0.00}&         &  9.5&  9.5&      &      & (8) \\
HAT-216-0003316   & 04 40 23.0 & +31 26 46 &       &       & M4V+M5V   &       &      &  54471.3745&   2.04813610&     &       &         & 15.2& 13.3&      &      & (9) \\
AP Tau            & 04 54 45.0 & +26 55 24 &       &       &           &       &      &  52500.1267&   0.97197470&     &       &         & 13.0&     &      &      & (5) \\
HAT-131-0026711   & 05 16 36.9 & +48 35 44 &       &       &           &       &      &  54497.1710&   0.66395310& 0.30&   0.10&         & 14.3&     &      &      & (9) \\
HAT-133-0002525   & 06 36 25.2 & +43 49 47 &       &       &           &  0.03 &      &  53632.4735&   1.59457150& 0.40&   0.16&   0.096 & 13.8&     &   4.8&      & (9) \\
V470 Cam          & 07 10 42.1 & +66 55 44 &   0.48&   0.13& sdB+M     &  0.04 &   6.3&  51822.7598&   0.09564665& {\it 1.00}&  {\it 0.20}&   0.015 & 14.7& 14.6&   1.2&   5.4& (10)\\
YY Gem            & 07 34 37.4 & +31 52 10 &   0.60&   0.60& dM1e      &  0.06 &   4.0&  52500.4573&   0.81428330& 0.55&   0.50&         & 10.6&  9.1&      &      & (11)\\
HAT-136-0003262   & 08 11 34.8 & +43 02 33 &       &       &           &       &      &  53770.8395&   0.64948470& 0.60&   0.65&   0.071 & 14.3&     &   3.4&      & (9) \\
GSC 1941 1746     & 08 25 51.9 & +24 27 04 &   0.56&   0.65& M2V+M2V   &       &   4.0&  53730.7303&   2.26560000& 0.90&   0.40&   0.453 & 12.9&     &   3.0&   1.3& (12)\\
CU Cnc            & 08 31 37.6 & +19 23 39 &   0.43&   0.40& M5Ve      &  0.03 &   5.1&  50208.5068&   2.77146800& 0.13&   0.11&   0.083 & 12.1& 11.4&   6.1&   0.8& (13)\\
NSVS 02502726     & 08 44 11.0 & +54 23 47 &   0.71&   0.35& K5V+M1V   &  0.04 &   4.3&  54497.5502&   0.55975500& 0.50&   0.35&   0.084 & 14.0& 13.4&   3.9&   1.1& (14)\\
GSC 2499 246      & 09 16 12.3 & +36 15 34 &   0.68&   0.73& M3V+M3V   &       &   3.6&  53456.6763&   0.36697100& 0.90&   0.60&   0.070 & 12.5&     &   1.0&   3.6&     \\
HAT-225-0003429   & 09 21 28.4 & +33 25 59 &       &       &           &  0.02 &      &  54534.1898&   0.42647590& 0.28&   0.27&   0.047 & 14.5&     &   6.6&      & (9) \\
BS UMa            & 11 25 41.0 & +42 34 50 &       &       & K9-M1V    &  0.10 &      &  52500.3477&   0.34950990& 0.41&   0.35&   0.059 & 12.5& 11.5&   2.0&      & (16)\\
HW Vir            & 12 44 20.2 &$-$08 40 17&   0.48&   0.14& sdB+M6-7  &       &   6.2&  52500.0560&   0.11671947& {\it 0.80}&  {\it 0.15}&   0.014 & 10.5&     &   0.2&  31.0& (17)\\
DE CVn            & 13 26 53.3 & +45 32 47 &   0.51&   0.41& M3V+DA    &       &   4.8&  52784.5533&   0.36413940& {\it 0.10}&  {\it 0.00}&         & 12.8& 12.2&      &      & (18)\\
NY Vir            & 13 38 48.1 &$-$02 01 49&   0.50&   0.15& sdB+M5    &       &   6.0&  52500.0594&   0.10101598& {\it 0.90}&  {\it 0.15}&   0.012 & 13.3& 13.5&   0.6&  10.0& (10)\\
NSVS 01031772     & 13 45 34.9 & +79 23 48 &   0.54&   0.50& M2V       &  0.02 &   4.4&  53456.6796&   0.36814140& 0.60&   0.60&   0.055 & 12.6& 11.0&   0.8&   5.2& (19)\\
HAT-145-0001586   & 13 45 13.2 & +46 18 40 &       &       &           &  0.01 &      &  53843.9266&   1.58752710& 0.65&   0.55&   0.064 & 14.3&     &   3.1&      & (9)\\
GK Boo            & 14 38 20.7 & +36 32 25 &       &       & K2V       &  0.04 &      &  52500.4350&   0.47777170& 0.92&   0.77&   0.091 & 10.6& 10.5&   0.5&      & (20)\\
GU Boo            & 15 21 54.8 & +33 56 09 &   0.60&   0.59& M0/M1.5   &  0.15 &   4.0&  52723.9811&   0.48871000& 0.90&   0.65&   0.064 & 13.1& 12.9&   1.2&   3.2& (21)\\
NSVS 07826147     & 15 33 49.4 & +37 59 28 &   0.38&   0.11& sdB+M5    &       &   7.2&  54524.0195&   0.16177042& {\it 1.35}&  {\it 0.20}&   0.016 & 13.0& 13.4&   0.4&  18.1& (15)\\
G179-55           & 15 47 27.4 & +45 07 51 &   0.26&   0.26& M4        &       &   7.0&  51232.8953&   3.55001840& 0.05&   0.06&         &     & 12.5&      &      &  (9)\\
NN Ser            & 15 52 56.1 & +12 54 45 &   0.54&   0.11& DAO1+M4   &       &   6.0&  52500.1209&   0.13008015& {\it $>$2} & {\it 0.00} &         & 16.7&     &      &      & (22)\\
HAT-192-0001841   & 16 12 16.7 & +41 13 51 &       &       &           &  0.02 &      &  53853.9056&   0.30873570& 0.62&   0.55&   0.037 & 14.0&     &   2.1&      &  (9)\\
CM Dra            & 16 34 20.4 & +57 09 44 &   0.23&   0.21& M4.5V     &  0.03 &   7.7&  52500.7177&   1.26839010& 0.75&   0.60&   0.038 & 12.9& 10.9&   9.9&   0.8& (23)\\
TrES-Her0-07621   & 16 50 20.7 & +46 39 01 &   0.49&   0.49& M3V+M3V   &       &   4.6&  53139.7495&   1.12079000& 0.11&   0.10&   0.090 & 15.5&     &  36.8&   0.1& (24)\\
HAT-196-0006238   & 17 58 59.3 & +35 55 12 &       &       &           &       &      &  53623.7449&   1.75834310&     &       &         & 14.9&     &      &      &  (9)\\
V924 Oph          & 18 33 28.3 & +07 07 51 &       &       &           &       &      &            &   0.35955400&     &       &         &     & 12.9&      &      &  (5)\\
OT Lyr            & 19 08 10.0 & +29 13 42 &       &       &           &       &      &  54222.4568&   0.47109500&     &       &         &     & 14.1&      &      &  (5)\\
FP Sge            & 20 14 45.8 & +19 36 49 &       &       &           &       &      &  52500.2947&   0.64200717&     &       &         &     & 14.0&      &      &  (5)\\
NSVS 14256825     & 20 20 00.4 & +04 37 56 &   0.46&   0.21& sdO+M2    &       &   5.9&  51288.9198&   0.11037410& {\it 0.75}&   {\it 0.20 }&   0.014 & 13.2& 13.3&   0.7&   7.9& (25)\\
FI Del            & 20 29 16.0 & +14 45 59 &       &       &           &       &      &     52500.3&   0.41592810&     &       &         &     & 14.6&      &      &  (5)\\
MR Del            & 20 31 13.5 & +05 13 08 &   0.69&   0.63& K0V       &  0.01 &   3.7&  52500.3087&   0.52169040& 0.33&   0.17&   0.073 & 11.0&  8.9&   1.4&   2.7& (26)\\
RX J2130.6+4710   & 21 30 18.5 & +47 10 07 &   0.55&   0.55& M4V+WD    &       &   4.2&  52785.6819&   0.52103563& {\it $>$2}  & {\it 0.00 } &         & 13.0&     &      &      & (27)\\
HS 2231+2441      & 22 34 21.5 & +24 56 57 &   0.30&   0.30& sdB+dM    &       &   6.3&            &   0.11058798&     &       &         & 14.0& 14.0&      &      & (28)\\
HAT-205-0007777   & 22 42 07.5 & +39 02 44 &       &       &           &       &      &  53146.0021&   1.11001010&     &       &         & 14.6&     &      &      & (9)\\
\hline
\end{tabular}
\end{center}
\flushleft{\footnotesize Explanation of columns: $\alpha_{2000}$, $\delta_{2000}$ - equatorial coordinates given for epoch and equinox 2000.0;
 $M_{1,2}$ - masses of the components; spectral classification; $A_{OCE}$ - observed O'Connell effect amplitude expressed as difference
 in maxima levels; $\Delta T$ - amplitude of the LITE for 1 Jupiter mass planet orbiting the binary on 10 years orbit; HJD$_{I}$, Period -
 ephemeris for the primary (deeper) minimum; $d_I$, $d_{II}$ - minima depth for the $I$ passband (for the $V$ passband typed in italics);
 $D_I$ - duration of the primary eclipse; $V,R$ - out-of-eclipse brightness of the binary; $\Delta t$ - theoretically achievable minimum
 precision using a 60cm telescope (equation~\ref{precision_min}) for the primary minima; $N_\sigma$ = $\Delta T$/$\Delta t$ detection 
 sensitivity of the binary. \\
 \smallskip
 References: (1) Zhang, Zhang \& Zhu (2010), (2) Law et al. (2012), (3) Cakirli \& {\.I}bano\u{g}lu (2010), (4) Dimitrov \& Kjurkchieva (2010),
 (5) Malkov et al. (2006), (6) Irwin et al. (2009), (7) Parsons et al. (2011), (8) Kundra \& Hric (2011), (9) Hartman et al. (2011),
 (10) Camurdan et al. (2012), (11) Torres \& Ribas (2002), (12) unknown, (13) Ribas (2003), (14) Cakirli et al. (2009),
 (15) For et al. (2010), (16) Wils et al. (2010), (17) Lee et al. (2009), (18) van den Besselaar et al. (2007),
 (19) L\'opez-Morales et al. (2006), (20) Zasche et al. (2012), (21) Windmiller et al. (2010), (22) Parsons et al. (2010),
 (23) Morales et al. (2009), (24) Creevey et al. (2005), (25) Wils et al. (2007), (26) Djura\v{s}evi\'c et al. (2011),
 (27) Maxted et al. (2004), (28) Ostensen et al. (2008)\\
 }
\end{table*}

Chances to discover a circumbinary substellar body depend primarily on three factors:(i) precision and number
of the minima which can be achieved; (ii) semi-amplitude of the LITE caused by the body; (iii) intrinsic variability
of the binary causing noise in minima timings.
The suitability of an object ($N_\sigma$ in Table~\ref{sampletab}) can be defined as the peak-to-peak amplitude of LITE
caused by such a body, $\Delta T$ divided by the theoretical precision of a single minimum timing, $\Delta t$, (see the
following subsections). Rather arbitrarily we included objects having $\Delta T/\Delta t > 1$. 

We consider most advantageous to focus on those binary systems with the richest information 
concerning their physical and orbital parameters (at least individual masses known). At the same time, we also plan 
to observe neglected systems to exclude objects with erroneous light-curve classification or identification.
For the best candidate eclipsing binaries with missing physical parameters we will strive to obtain dedicated 
spectroscopy at the Rozhen observatory (Bulgaria) or applying for observing time at major observatories.

\subsection{Expected precision of the minima timings}
\label{precision}

The goal of the project is to determine the times of the minima with the highest possible precision. For the
systems with triangular shape of the minima (i.e., binaries with partial eclipses and non-degenerate 
components\footnote{In the case of eclipses of degenerate components, where egress and ingress last 
typically from couple of seconds to a minute, the time resolution of the photometry defines the timing 
precision.}) the brightness uncertainty (in magnitudes) of a single observational point, $\sigma$, is 
clearly related to its time uncertainty, $\delta t$ by the slope of the minimum branch. Then the minimum 
uncertainty, $\Delta t$, can be estimated as:

\begin{equation}
\label{eqn1}
\Delta t = \frac{D \sigma}{2 d \sqrt{N}},
\end{equation}

where $d$ is the depth (in magnitudes) and $D$ duration of the minimum, and $N$ number of observational points during the
eclipse. The above relation
shows that the precision of the minimum time increases with the number of datapoints taken in the minimum
and their precision, which mostly relates to the diameter of the telescope. The shape of minimum also
affects the precision of the timing - deep and at the same time narrow minima provide the best 
precision\footnote{Doyle et al. (1998) suggested exact but substantially more complicated approach
taking into account exact shape of the minimum and sampling of the observations.}.

The expected precision of the minima (as given by the shot noise) clearly depends on the size of the
telescope and exposure time. Uncertainty of an individual observing point is inversely proportional 
to the product of collecting area of the telescope and exposure time, $E$ (assuming no read-out nor scintillation
noise). If $A$ is the aperture or the diameter of the telescope used, then:

\begin{equation}
\label{eqn2}
\sigma \propto \frac{1}{\sqrt{\pi \left(\frac{A}{2} \right)^2 E}}.
\end{equation}

Assuming no read-out overheads (then $N = D/E$) combining (\ref{eqn1}) and 
(\ref{eqn2}) one can find that:

\begin{equation}
\label{eqn3}
\Delta t = C \frac{\sqrt{D}}{\sqrt{\pi} A d}.
\end{equation}

Proportionality factor, $C$, can be quantified counting the photons coming from the source taking into
account the telescope throughput, filter transparency, quantum efficiency of the detector and atmospheric 
extinction. If the average brightness of the target during the eclipse is $m_\lambda$, and it is observed at the 
air mass $X$ at (the wavelength-dependent) extinction coefficient $\kappa_\lambda$ then (see Sections 1.7, 
1.8 and 2.5 of Henden \& Kaitchuck, 1982):

\begin{equation}
\label{precision_min}
\Delta t = \frac{1}{\sqrt{\tau F_\lambda}} 10^{0.2(m_\lambda+X\kappa_\lambda)} \frac{\sqrt{D}}{\sqrt{\pi} A d}.
\end{equation}

The constant $\tau$ is the throughput of the observing system\footnote{$\tau \in (0,1)$ depends on the quantum efficiency
of the CCD chip, the absolute transparency of the filter used, the reflection losses on the telescope mirrors etc.}, and
$F_\lambda$ is number of photons from a $m_\lambda$ = 0 star per square meter and second outside Earth atmosphere recorded
through the filter used. Preliminary observations from the 60-cm telescope equipped with the MI G4-9000 CCD camera at 
Star\'a Lesn\'a show that from a $R$ = 14-mag star, we record 250 photons in one second in the $R$ passband at air mass $X$ = 1 and
$\kappa_R$ = 0.2, which translates to $\tau \approx 0.16$.

Because of several other sources of the noise and non-negligible read-out times (8 sec in the case of
the 2x2 binning read-out with the MI G4-9000 CCD camera) the minima uncertainties as estimated by relation
(\ref{precision_min}) and listed in Table~\ref{sampletab} should be regarded as the theoretical limits.
The estimates, nevertheless, immediately show the suitability of the selected targets for the present
project.


\subsection{LITE amplitude}

It can easily be shown that the full amplitude (Max - Min or peak-to-peak) of the expected LITE changes
caused by another body orbiting a binary on the edge-on ($i \sim 90\degr$) circular orbit ($e \sim 0$) is:

\begin{equation}
\label{lite_amp}
\Delta T \approx \frac{2 M_3 G^{1/3}}{c} \left[\frac{P_3}{2\pi (M_1 + M_2)} \right]^{2/3},
\end{equation}

where $M_1, M_2, M_3$ are masses of the components, $G$ is gravitational constant, $c$ is speed of light,
and $P_3$ is orbital period of the third (substellar) component. Eqn.~\ref{lite_amp} shows that the
semi-amplitude of the LITE changes is proportional to the mass of the third component and nearly proportional
to its orbital period. The total mass of the underlying binary is also important: having 8 times more
massive inner binary decreases the LITE amplitude four times. The advantage of low-mass binaries is,
on the other hand, offset by their surface activity causing noise or spurious periodicities in the timing data.

Table~\ref{sampletab} quotes, whenever $M_1 + M_2$ is available, the estimated full amplitude,
$\Delta T$, in the case of a Jupiter-mass planet orbiting the EB in 10 years. The full (peak-to-peak)
amplitude ranges between 4 and 8 seconds. Detecting a Jupiter-mass planet on a shorter orbit is, probably,
beyond the expected precision of the timing data: shortening the planet's orbital period 8 times to 1.5 years,
would decrease the full amplitude four times to about 1-2 seconds for our targets.

\subsection{Intrinsic variability of the binary star}
\label{variability}

In addition to the timing errors caused by precision and accuracy of the data acquisition, the minima times
are affected by intrinsic variability of the binaries. The dominant cause of the intrinsic variability (and
enhanced scatter) in the (O-C) diagrams are spots on the active late-type K or M dwarf components.
The level of the activity is the highest in the tidally bound close binaries, where magnetic braking could
not slow down the rotation. This is the case for systems with orbital periods shorter than about 0.5-1 day.
The level of the activity and its effect on the timing varies from system to system. In the most active
systems the amplitude of the photometric variations reaches 0.2-0.3 mag (e.g. in DK~CVn, Terrell et al., 2005).

Dark photospheric spots seen in the majority of the late-type systems cause LC asymmetries (O'Connell
effect) and out-of-eclipse photometric wave(s). A spot seen by the observer close to or during the minimum
of light shifts it from the spectroscopic conjunction. A single dark/hot spot not eclipsed during
the minimum causes a small linear trend across it. This shifts the observed instant of the minimum from the
spectroscopic conjunction. If the maximum following the primary minimum is the brighter/fainter one, then
the observed instant of minimum is observed earlier/later than the instant of the spectroscopic conjunction.

The maximum time shift caused by a single starspot can be estimated as follows:

\begin{equation}
\Delta t = \frac{2A_{OCE} D^2}{dP},
\end{equation}

where $A_{OCE}$ is {\it observed\footnote{Most easily estimated as the maxima difference.}} peak-to-peak amplitude 
(see Table~\ref{sampletab}) of the out-of-eclipse photometric wave, $D$ is the duration of minimum, $P$ is the 
orbital period, and $d$ is the depth of eclipse. The equation shows that the effect of spots is proportional 
to the photometric wave amplitude while inversely proportional to the depth of minima. The magnitude of the 
shift also depends on the method used to determine the minimum instant.

To check for the spot effects which might shift the minimum time from the spectroscopic conjunction,
it is advisable to cover parts of the LC just before ingress and just after the egress from the minimum.
The difference in the shoulders levels indicates the amplitude of the photometric wave. Another
indicator of a spurious shift caused by surface maculation is the minimum asymmetry. This
can be checked by the minimum bisectors slant (similar to a technique used in RV exoplanet searches).
The bisector method provides best minimum instant estimate as well as realistic errors (see
Covino et al., 2004). 

When a single photospheric spot causes a trend across the minimum its effect can be rectified using the
fitting formula~ (\ref{model}) which will be discussed in the following section. The presence of several 
spots or spots being eclipsed during the minimum complicates the rectification of the effect. The 
maximum O'Connell effect is listed in Table~\ref{sampletab} as OCE. It is pertinent to the $I$ passband LC.

Binaries composed of a hot sdB primary and a M dwarf secondary do not show pronounced ($>$0.02 mag)
LC asymmetries. This results from the dominant contribution of the hot subdwarf to the total light of
the system. The secondary minimum occurs just due to the eclipse of the irradiated hemisphere
of the secondary by the hot subdwarf\footnote{The true light contribution of the secondary can be 
seen from the difference in the level of shoulders and that of the mid secondary eclipse, see Fig.~\ref{lcsample}.}.

Pulsations of the sdB components, in e.g. NY~Vir, pose less serious complication. With the pulsation
amplitude not exceeding 0.01 mag and the minimum depth of 0.5-1 mag the shift of the minimum is negligible
(see Kilkenny, 2011). To exclude multi-periodic variability of the studied EBs a detailed period analysis
of each object will be performed prior to the determination of the minima times. In the case of the
positive detection of pulsational variability the LCs will be iteratively pre-whitened.

\section{Data reduction process}
\label{reduction}

\subsection{CCD frames reduction}
\label{ccd}

The reduction of the CCD frames and aperture photometry will be done using scripts written under the IRAF
package\footnote{\tt http://iraf.noao.edu/}. The same scripts and approach will
be used at all participating institutions.

In the first step, the master dark and flat-field frames will be produced for all exposure times, filters and
CCD temperatures. To reduce effects of scattered light usually seen in sky flatfields the master flat fields
will be box-car average divided to remove low-frequency variations while pixel-to-pixel sensitivity differences
will be preserved. The illumination effects due to vignetting of the CCD chip will be minimized by keeping the
EB and its comparison star(s) at approximately the same pixel coordinates (most of the instruments are
autoguided).

In the next step, the raw CCD frames will be dark and flatfield corrected. Then the WCS system will be
determined using the GSC 2.3.3 online catalogue for
reference\footnote{\tt http://gsss.stsci.edu/webservices/GSC2/\\GSC2WebForm.aspx}. Finally, aperture photometry of
the target and all stars with a brightness within $\pm$1 mag off target brightness will be performed. The instrumental
magnitudes will be determined for several apertures appropriate for the seeing range at the site. All data of a given
target from one instrument would be analyzed at the same time to see variability of {\it all} measured stars on the
frames and to select all stable stars to produce an artificial comparison star (see e.g., Broeg et al., 2005).

The resulting LCs will be available on the DWARF project pages\footnote{\tt http://www.ta3.sk/$\sim$pribulla/Dwarfs/} 
for subsequent LC analysis (e.g., determination of photometric elements or detection of pulsations, flares) or 
redetermination of the minima times.

There are two major systematic complications which can cause spurious shifts in the minima timing: (i) shutter effect
during short exposures (ii) non-linearity effects. While the former complication caused by finite speed of the
shutter opening can be avoided by using a nearby comparison and avoiding exposures below 10 seconds, the latter
effect is harder to quantify. The non-linearity effect, fortunately, does not affect the timing in the case of 
symmetric minima but can reach a few per cent even well below saturation levels. 

\subsection{Reference time for the data}
\label{time}

Because the goal of the project is going to be accomplished by the accurate (at a level of a few seconds) timing of
the LC extrema (in our case minima), it is crucial to regularly synchronize the computer clock with
the ntp servers to provide the system time within 1 second off the UTC. Another systematic time shift is
connected with the shutter delay: the FITS header gives the instant when the imaging started but it takes
a few tenths of a second for the shutter to fully open. This effect will have to be quantified to avoid systematic
minima shifts.\footnote{Tests at Astronomical Station Vidojevica/Serbia show that the shutter delay of
Alta Apogee E47 CCD camera amount to 0.42 sec.}

UTC (or Coordinated Universal Time) is based on the atomic clocks but it is never allowed to differ from UT1
(based on the rotation of Earth) by more than 0.9 seconds. Therefore a leap second has to be included from
time to time (next on June 30, 2012). The UTC, therefore, is discontinuous and not the best to use in timing
analysis. It is fully correct to use Barycentric Dynamical Time (TDB) which takes into account relativity --
the fact that moving clocks tick at different rates. TDB is a truly uniform time, as we would measure it on
Earth if it were not moving around the Sun or being pulled by the Moon and other celestial bodies.

Using HJD (Heliocentric Julian Date) correction is also not sufficient: the mass center of the Sun revolves around
the Solar system barycenter. This causes errors which can reach up to 4 seconds (mostly defined by the orbits of
Jupiter and Saturn). Therefore it is best to relate time to the Solar System Barycenter.

Hence we will use Barycentric Julian Dates in Barycentric Dynamical Time (BJD-TDB) \footnote{BJD-TDB differs
from the BJD in Coordinated Universal Time (BJD-UTC) by a systematic 32.184 + $N$ seconds, where $N$ is the
number of leap seconds that have elapsed since 1961 ($N$ = 34 as of Jan 1st, 2009).}.

To allow for possible corrections or improvements of UTC to BJD-TDB we will also list times of minima
in JD based on UTC (without HJD or BJD correction added).

\subsection{Minima determination}
\label{minima}

The most widespread approach to obtain instants of minima of EBs is to use the Kwee \& van Woerden (1956) method.
From our experience, the errors estimated using their formula (14) are often unrealistically small. The real
uncertainties are often dominated by systematic errors.

The LCs are affected (at the 0.01 mag level) by the red noise (spurious shifts and trends) caused mostly
by the atmosphere transparency changes, lack of autoguiding in some of the telescopes and the second-order
extinction (Forbes effect). The last effect causes photometry errors even at the same air mass because of
differences of spectral types of the target and comparison star(s) (see e.g., Pakstiene \& Solheim, 2003). The
second-order extinction will be not negligible especially for EBs of very late spectral type when no appropriate
comparison star would be found in the field.

To reduce effects of scattered light on sky-flats (centers are most illuminated usually) a box-car averaging
will be used (see Section \ref{ccd} and Freundling et al., 2007). This would minimize night-to-night 
offsets of the LCs.

Our photometry could possibly be improved by using the algorithm based on principal component analysis proposed by
Tamuz, Mazeh \& Zucker (2005). Unfortunately, the CCD frames will be obtained at several observatories with
different setups and even different orientation of the field. Systematic errors in minima positions would be
partially removed by the fitting technique proposed below.

For each EB, the fitting templates will be prepared to obtain the instant of conjunction (minimum) for
any sufficiently long photometric sequence. Such a way, we will use not only the minima but also other LC
segments where the brightness sufficiently changes. The template LC will be produced as the average obtained
over the whole campaign.

For EBs with components of widely different temperatures (e.g., sdB + dM systems, or WD + K systems) the amplitude
and (relative) depth of minima strongly depends on the wavelength of the observation. Due to the differences in
filter transparencies and wavelength response of detectors, we will form a template LC for each filter separately
and the fitting LC will be scaled to match the observations. From our former observing experience, we could
also note small nightly shifts of the LCs observed even with the same instrument. Sometimes the LC shows slight
but systematic slopes. These slopes are, very probably, caused by scattered light combined with drifting of the
targets on the CCD due to imperfect tracking of the telescope.

To obtain good fits of the template $T(x)$ to the observed LCs (and accurate timings), we constructed the following
fitting function (see Pribulla et al. 2008):

\begin{equation}
\label{model}
F(x) = A + Bx + CT(x - D),
\end{equation}

\noindent which would allow for shifting, scaling and 'slanting' of the template LC. Fixing of the parameters will
be judged according to the appearance of individual LCs, e.g. in the case that only one branch was observed
the vertical shift ($A$) would be fixed to zero.

\subsection{Minima uncertainties}
\label{uncertainties}

As we discussed above, the errors estimated using the classical method for minima determination are usually unrealistically
small. To get more realistic estimates and to see the correlations of the parameters, Monte Carlo simulations
will be used. To preserve the original shape and scatter of the data, the fitting function $F(x)$ in the instants
of real observations will be replicated adding the Gaussian-distributed random noise. The standard deviation of the
added noise will correspond to the standard deviation of the original data with respect to the original fit. Preliminary
tests show that about 2000 replications of the LC are sufficient to arrive at the errors\footnote{The standard deviation of given
parameters will be determined from the width of its Gaussian distribution profile.} and correlations of the
parameters.

We also considered the bootstrapping of the data (see Press et al. 1993; Efron \& Tibshirani, 1993). Because of the small
number of datapoints on branches (especially for short-period and WD systems) the randomly re-sampled LCs could skip
the crucial phases defining the minimum time and its precision.


\section{The timing analysis and its limitations}
\label{OC}

If an unseen third component revolves an EB, the residuals with respect to a linear (or quadratic) ephemeris will
show wavelike behavior in the (O-C) diagram because of the LITE.

Although all our targets are well detached systems (where the mass transfer cannot occur), the orbital period can
continuously decrease for two reasons (i) the magnetic braking in the case of systems with late-type components
(ii) radiation of gravity waves in the case of the systems with shortest orbital periods ($<$ 0.1 day).

(O-C) curve (due only to the inner binary) can be described by a linear (constant period) or quadratic ephemeris 
(linear period variation) and if a third body is orbiting the inner binary adding the LITE effect, the times of 
the minima can be computed as follows:

\begin{eqnarray}
Min~I = JD_0 + P \times E + Q \times E^2 + \nonumber \\
+ \frac{a_{12} \sin~i}{c}\left[\frac{1 - e^2}{1 + e\cos~\nu} \sin(\nu + \omega) +
e \sin~\omega \right],
\end{eqnarray}

\noindent where $a_{12}$~sin~$i$ is the projected semi-major axis (inclination cannot be derived from the LITE alone), 
$e$ is the eccentricity, $\omega$ is the longitude of the periastron, $\nu$ is the true anomaly of the EB 
orbit around the common center of the mass of the whole system. $JD_0 + P \times E + Q \times E^2$ is the 
quadratic ephemeris of the minima of the EB and $c$ is the velocity of light. The parameter $Q$ is the coefficient 
of the quadratic term and gives the rate of the orbital period change of the EB. If $a$ is the semi-major axis 
of the third (substellar) companion's orbit around the binary's mass center, then:

\begin{equation}
a_{12}=\frac{a M_3}{M_1+M_2+M_3}.
\end{equation}

\noindent To obtain the optimal fit and corresponding elements ($JD_0$, $P$, $Q$, $a$~sin~$i$, $e$, $\omega$, and also 
epoch of periastron passage, $T_0$, and the period of the orbit of three-body system, $P_3$) of the LITE orbit 
including errors, we use the differential corrections method (see Irwin 1959).

The orbital elements determined above enable to estimate the mass of the unseen companion. In the case we know
the masses of the binary star components $M_1, M_2$ (or their sum), we can derive the mass function of the
third-body:

\begin{eqnarray}
f(M_3) = \frac{(M_3 \sin~i)^3}{(M_1 + M_2 +M_3)^2} = \nonumber \\
= \frac{4\pi^2}{G}\frac{(a_{12} \sin~i)^3}{P_3^2} = \frac{4\pi^2}{G} \frac{A^3 c^3}{P_3^2}
\end{eqnarray}

\noindent where $A$ is the semi-amplitude of LITE (in time units), $i$ is the inclination of the orbital 
plane of the third body, $M_1$, $M_2$ are the masses of the eclipsing binary components, and $M_3$ is the mass 
of the third (substellar) companion, $P_3$ is the outer orbital period, and $G$ is the gravitational constant. 

Assuming that the
mass of the substellar body is negligible compared to the total mass of the binary, $M_3 \ll M_1 + M_2$, its
mass can be directly found as:

\begin{equation}
M_3^3 \sin^3 i \approx \frac{4\pi^2(M_1+M_2)^2}{GP_3^2} A^3 c^3
\end{equation}

The analysis of the selected EBs timings will be performed in three steps: (i) period search in the (O-C) residuals
with respect to a linear or quadratic ephemeris,
(ii) fitting LITE orbits to most promising orbital periods. Orbital periods longer than the time span of the data
will not be considered. A major problem in the timing analysis is matching our uncertainties with those listed
for the published timings. Moreover, the minima uncertainties are often not given. This will complicate
the relative weighting of the datapoints and cause additional uncertainty of the results.

In the case of K or M dwarf systems both primary and secondary minima would be used. There is no system in our
sample showing elliptic orbit. Hence, both types of minima can be simultaneously analyzed. Eventually, we can 
check the results of the (O-C) curves obtained with primary and secondary minima only before merging both times
of minima in a unique curve.

The observed cyclic variations can also be caused by the magnetic-orbital moment coupling mechanism of
Applegate (1992). The mechanism causes periodic variations of the orbital period, out-of-eclipse brightness
and color. The brightness, color and period changes show the same periodicity. This fact can be used to
tell apart the LITE from Applegate's mechanism.

\section{Observing network and first observations}
\label{network}

The targets  will be observed at several observatories using 35-120cm telescopes equipped mostly with low-end
CCD cameras (see Table~\ref{telescopes}). All observatories are in the Northern hemisphere. Hence we limit our
targets to those north of DEC = $-10$\degr. The longitude spread of the observatories is much larger. This is
important because of several EBs with orbital periods longer than one day and for the systems
with orbital period being close to integer multiple of one day (e.g., AP~Tau or HAT-216-0003316).

\begin{table*}
\begin{scriptsize}
\caption{Telescope network (status as of April 22, 2012)
\label{telescopes}}
\begin{center}
\begin{tabular}{lrrlrlccl}
\hline
\hline
Observatory                 &    Long.  & Lat.   & Telescope     & Aperture  & Camera              & CCD size             & FoV             & Ref.  \\
                            &   [deg.]  & [deg.] &               & [cm]      &                     &                      & [arcmin]        &       \\
\hline
SOAO/Korea                  & 128.4E    & 36.9N  & Cassegrain    & 60        & E2V CCD42-40        & 2048 $\times$ 2048   & 18 $\times$ 18  & (1)   \\
Terskol/Russia              &  42.5E    & 43.3N  & Cassegrain    & 60        & Pixel Vision        & 1024 $\times$ 1024   & 10$\times$10    &       \\
                            &           &        & Schmidt-Cass. & 35        & SBIG STL-1001       & 1024 $\times$ 1024   & 24$\times$24    &       \\
                            &           &        & Schmidt-Cass. & 29        & S3C                 & 1024 $\times$ 1024   & 28$\times$28    &       \\
OMU/Turkey                  &  36.2E    & 41.4N  & Schmidt-Cass. & 35        & STL-4020M           & 2048 $\times$ 2048   & 15 $\times$ 15  &       \\
Ankara/Turkey               &  32.8E    & 39.8N  & Schmidt-Cass. & 40        & Apogee Alta U-47    & 1024 $\times$ 1024   & 11 $\times$ 11  & (2)   \\
Kottamia/Egypt              &  31.8E    & 29.9N  & Cassegrain    & 188       & EEV CCD 42-40       & 2048 $\times$ 2048   &  2.7$\times$2.7 & (3)   \\
MAO NASU/Ukraine            &  30.5E    & 50.4N  & Cassegrain    & 70        & SBIG STL-1001       & 1024 $\times$ 1024   & 24$\times$24    &       \\
Lesniki/Ukraine             &  30.5E    & 50.3N  & Schmidt-Cass. & 35        & Rolera MGi          &  512 $\times$ 512    &  7$\times$7     &       \\
ITAP/Turkey                 &  28.3E    & 36.7N  & Schmidt-Cass. & 35        & SBIG ST10 XME       & 2184 $\times$ 1472   & 14 $\times$ 10  & (4)   \\
Ege/Turkey                  &  27.1E    & 38.4N  & Schmidt-Cass. & 40        & Apogee CCD47-10     & 2048 $\times$ 2048   & 20 $\times$ 20  & (5)   \\
Rozhen/Bulgaria             &  24.7E    & 41.7N  & Cassegrain    & 60        & FLI ProLine 09000   & 3056 $\times$ 3056   & 17 $\times$ 17  & (6)   \\
                            &           &        & Schmidt       & 50/70     & FLI ProLine 16803   & 4096 $\times$ 4096   & 73 $\times$ 73  &       \\
                            &           &        & Rit.-Chret.   & 200       & Vers Array 1300B    & 1340 $\times$ 1300   & 5.7$\times$ 5.7 &       \\
Feleacu/Romania             &  23.6E    & 46.7N  & Schmidt-Cass. & 40        & SBIG STL-6303E      & 3072 $\times$ 2048   & 23 $\times$ 16  & (7)   \\
Kolonica                    &  22.3E    & 48.9N  & Cassegrain    & 100       & FLI PL1001E         & 1024 $\times$ 1024   & 10 $\times$ 10  & (8)   \\
Slovakia                    &           &        & Schmidt-Cass. & 35        & MI G2-1600          & 1536 $\times$ 1024   & 12 $\times$  8  &       \\
                            &           &        & Schmidt-Cass. & 50        & MI G4-16000         & 4096 $\times$ 4096   & 31 $\times$ 31  &       \\
Patras/Greece               &  21.7E    & 38.3N  & Schmidt-Cass. & 35        & SBIG ST10 XME       & 2184 $\times$ 1472   & 20 $\times$ 14  &       \\
Astron. Station Vidojevica  &  21.5E    & 43.1N  & Cassegrain    & 60        & Apogee Alta U-42    & 2048 $\times$ 2048   & 16 $\times$ 16  & (9)   \\
Serbia                      &           &        & Cassegrain    & 60        & Apogee Alta U-47    & 1024 $\times$ 1024   & 7.6$\times$ 7.6 & (10)  \\
Roztoky/Slovakia            &  21.5E    & 49.4N  & Cassegrain    & 40        & MI G2-1600          & 1536 $\times$ 1024   & 12 $\times$ 8   &       \\
Star\'a Lesn\'a             &  20.3E    & 49.2N  & Newton        & 50        & SBIG ST10 XME       & 2184 $\times$ 1472   & 20 $\times$ 14  & (11)  \\
Slovakia                    &           &        & Cassegrain    & 60        & MI G4-9000          & 3056 $\times$ 3056   & 17 $\times$ 17  &       \\
Szeged/Hungary              &  20.2E    & 46.2N  & Newton        & 40        & SBIG ST7            & 765  $\times$ 510    &  17$\times$ 11  &       \\
Toru\'n/Poland              &  18.6E    & 53.1N  & Cassegrain    & 60        & SBIG STL-1001       & 1024 $\times$ 1024   & 12 $\times$ 12  &       \\
Brno/Czech Rep.             &  16.6E    & 49.2N  & Newton        & 62        & SBIG ST8            & 1530 $\times$ 1020   & 17 $\times$ 11  &       \\
                            &           &        & Schmidt-Cass. & 35        & G2-4000             & 2056 $\times$ 2062   &                 &       \\
Hvar/Croatia                &  16.4E    & 43.2N  & Cassegrain    & 100       & Apogee Alta U-47    & 2048 $\times$ 2048   &  8$\times$8     & (12)  \\
Graz/Austria                &  15.5E    & 47.1N  & Astro\_Topar  & 30        & SBIG STL11000M      & 4008 $\times$ 2672   & 16 $\times$ 11  &       \\
                            &           &        & Cassegrain    & 50        & SBIG ST-2000XM      & 1600 $\times$ 1200   &  9 $\times$ 7   &       \\
Catania/Italy               &  15.0E    & 37.7N  & Cassegrain    & 91        & KAF1001E            & 1024 $\times$ 1024   & 12.5$\times$ 12.5 &(13) \\
                            &           &        & Rit.-Chret.   & 80        & Apogee  U9000       & 3040 $\times$ 3040   &  17 (diameter)  & (14)  \\
Prague/Czech Rep.           &  14.4E    & 50.1N  & Schmidt-Cass. & 40        & SBIG ST10 XME       & 2184 $\times$ 1472   & 24 $\times$ 16  & (15)  \\
TLS/Germany                 &  11.7E    & 51.0N  & Schmidt       & 30        & Apogee AP-16        & 4096 $\times$ 4096   &132$\times$132   & (16)  \\
Jena                        &  11.5E    & 50.9N  & Schmidt       & 60/90     & E2V CCD42-10        & 2048 $\times$ 2048   & 53 $\times$ 53  & (17)  \\
Germany                     &           &        & Cassegrain    & 25        & E2V CCD47-10        & 1056 $\times$ 1027   & 21 $\times$ 20  & (18)  \\
Kirchheim/Germany           &  11.0E    & 50.9N  & Rit.-Chret.   & 60        & SBIG STL-6303E      & 3072 $\times$ 2048   &  71$\times$52   & (19)  \\
Herges-Hallenberg/Germany   &  10.6E    & 50.7N  & Cassegrain    & 20        & MI G2-1600          & 1536 $\times$ 1024   &  48$\times$32   & (20)  \\
Trebur/Germany              &   8.4E    & 49.9N  & Cassegrain    & 120       & SBIG STL-6303E      & 3072 $\times$ 2048   & 10 $\times$ 7   & (21)  \\
LOAO/USA                    & 110.7W    & 32.4N  & Cassegrain    & 100       & ARC 4K CCD          & 4096 $\times$ 4096   & 28 $\times$ 28  & (22)  \\
\hline
\hline
\end{tabular}
\end{center}
\end{scriptsize}

\flushleft{\footnotesize Notes: 
(1)~Lee, Kim \& Koch (2007);
(2)~{\tt http://rasathane.ankara.edu.tr/en/tools\_info.php?id=2};
(3) field of view given in the Cassegrain focus, see Azzam et al. (2010);
(4)~{\tt http://itap-tthv.org/astro/gozlemevi.html};
(5)~{\tt http://astronomy.sci.ege.edu.tr/ASTRO-WEB/TR2/};
(6)~{\tt http://www.nao-rozhen.org/telescopes/fr\_en.htm} the 60cm Cassegrain has 26$\times$26 arcmin field of view with the focal reducer;
(7)~Turcu et al. (2011);
(8)~{\tt http://www.astrokolonica.sk/joomla15/index.php/sk/instruments.html};
(9)~{\tt http://belissima.aob.rs/}; 
(10)~with 0.6$\times$ focal reducer;
(11)~Pribulla \& Chochol (2003);
(12)~{\tt http://oh.geof.unizg.hr/index.php/en/instruments/austro-croatian-telescope};
(13)~{\tt http://sln.oact.inaf.it/index.php/it/telescopio-91-cm/strumentazione/camera-ccd.html};
(14)~{\tt http://sln.oact.inaf.it/index.php/it/telescopio-apt2.html}; 
(15)~\v{S}tef\'anik Observatory {\tt http://www.observatory.cz/};
(16)~{\tt http://www.tls-tautenburg.de/tls\_d.php?category=test\_d};
(17)~Mugrauer \& Berthold (2010), STK - Schmidt Telescope Camera;
(18) CTK - Cassegrain Telescope Camera);
(19) Volkssternwarte Kirchheim;
(20) private observatory at Herges-Hallenberg; 
(21) http://homepages-fb.thm.de/jomo/t1t.htm;
(22)~Lee et al. (2012).
}
\end{table*}

To get the best S/N it is advisable to use the $R$ or $I$ filter for M or K EBs (see Section~\ref{sample}),
and the $V$ filter (or $B$ in the case of back-illuminated CCDs) for the systems with sdB or WD components
\footnote{The eclipse depth in WD systems is quickly increasing to the shorter wavelength range.}. Several
faint or short-period objects will be observed without filter to provide more light. Using two or more
filters would decrease the cadency of the photometry and decrease the duty-cycle because of filter change overheads
(this is the case for simple CCD cameras which do not enable simultaneous observing in several passbands).
For short-period systems it is advisable to cover both minima shoulders to see the LC asymmetry caused by the
photospheric spots.

\begin{figure*}
\includegraphics[width=170mm,clip=]{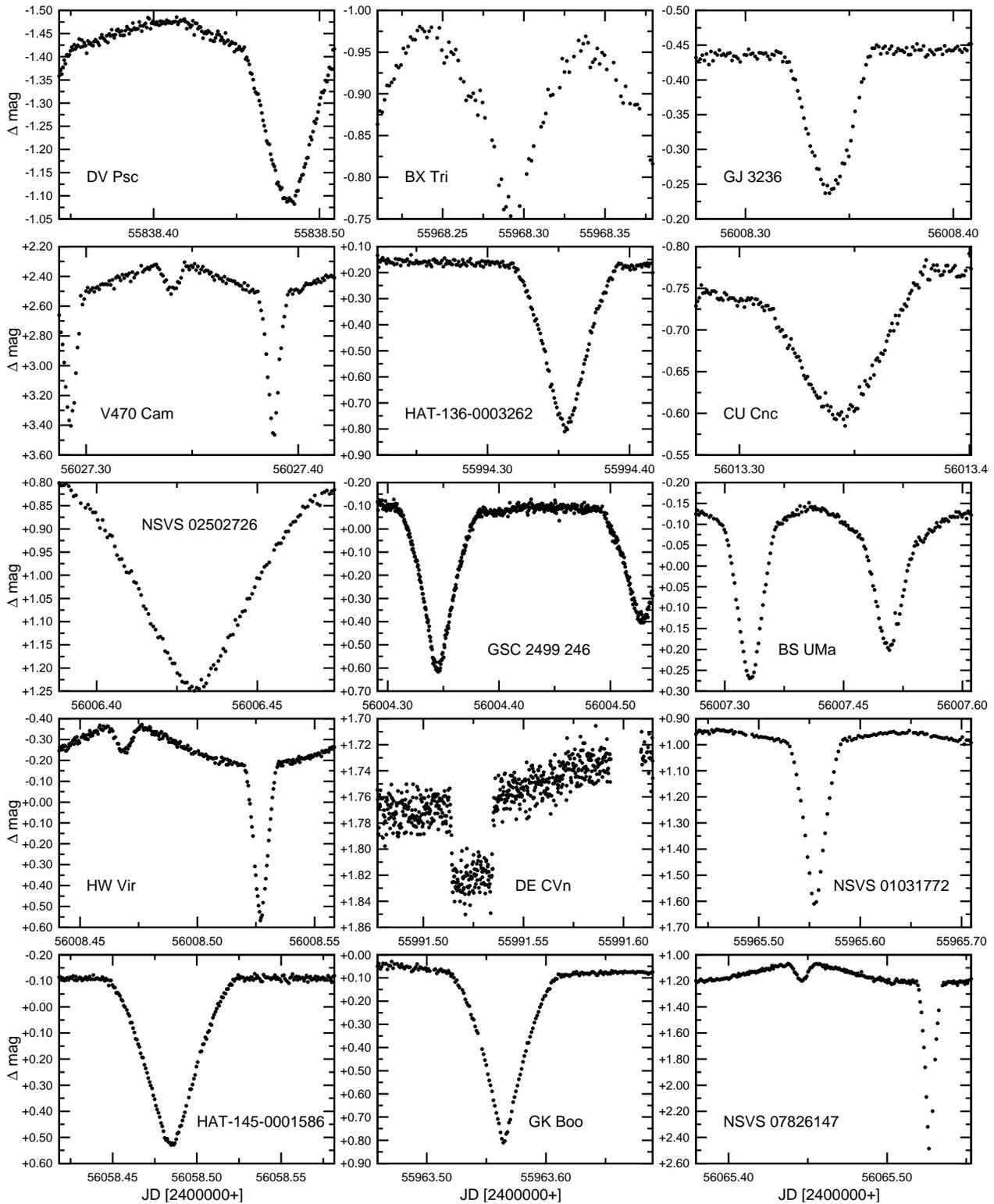}
\caption{A sample of LCs obtained at the Star\'a Lesn\'a observatory using either 50cm Newtonian or 60cm Cassegrain
         telescopes. The Julian dates are geocentric. LCs of most objects were obtained in the $I$ passband
         (except V470~Cam, HW~Vir, DE~CVn and NSVS~07826147 observed in the $V$ passband). 
\label{lcsample}}
\end{figure*}

For most binaries with M and K dwarf components both minima (primary and secondary) are useful for
the timing analysis. In the case of sdB EBs or systems with a WD component only the primary minimum will be
observed. The secondary eclipse is too shallow (or hardly to be detected) to provide sufficient timing precision
in most of those systems.

In addition to the observations focused on the exact timing, we will perform (much less extensive) multi-color
$UBVRI$ photometry of the same fields to find the best comparison stars (to minimize the second-order
extinction effects). The observations will also be focused to check possible out-of-eclipse and color brightness
variations which would indicate the mechanism causing cyclic variations of the orbital period proposed by
Applegate (1992).

The medium to high-resolution spectroscopy at the 2m telescope at the Rozhen observatory in
Bulgaria\footnote{\tt http://www.nao-rozhen.org/index\_bg.htm} will be performed for objects without any
spectroscopy to infer the nature and spectral type of the components and to exclude systems with stellar third
or multiple components.

Systematic CCD observations of the targets (Table~\ref{sampletab}) started at the Star\'a Lesn\'a Observatory in
February 2012 (see a sample of LCs obtained during the tests of the observing facilities at Star\'a Lesn\'a 
in Fig.~\ref{lcsample}). The precision of the data is good but still hardly approaching the theoretical limits given in the
Table~\ref{sampletab}. The best error estimates of 4 primary minima determined from the preliminary
reduction (differential photometry with respect to one comparison) of the CCD photometry of NY~Vir
(March 16, 24, and 26, 2012) range between 1.7 and 3.5 seconds while the theoretically estimated limit is
0.6 seconds. This results from the red noise in the data (no autoguider available), possible variability
of the comparison star (GSC~4966-00559), and read-out overheads.

\begin{figure}
\includegraphics[width=80mm,clip=]{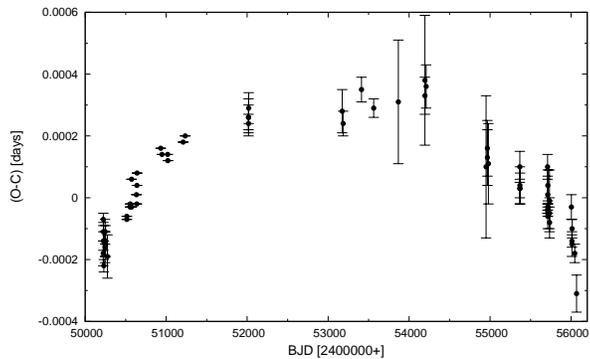}
\caption{{\it O-C} diagram for the sdB + M dwarf binary NY~Vir. Six last points grouped around
         BJD 2\,456\,000 were determined from the CCD photometry obtained with the 60cm Cassegrain
         telescope at Star\'a Lesn\'a (see Table~\ref{telescopes}). The (O-C) values were computed
         using linear ephemeris of Qian et al. (2012), Min I = BJD$2\,450\,223.362213(8) + 0.1010159673(2)\times E$.
\label{nyvir}}
\end{figure}

The (O-C) diagram of NY~Vir using all published (Kilkenny et al., 1998, 2000; H\"ubscher, Paschke \& Walker, 2006;
Kilkenny, 2011; Zasche et al., 2011; Qian et al., 2012; Camurdan et al., 2012) and 6 new minima timings
is presented in Fig.~\ref{nyvir}. Our data are consistent with the overall period decrease in the
system.

\section{Conclusion}

The presented project is aimed at the detection of circumbinary extrasolar planets and brown dwarfs using
minima timing variability of carefully selected EBs. Unlike more widespread techniques (RV or transit
searches) to detect extrasolar planets, the minima timing does not require high-end and costly astronomical
instrumentation. Precise photometric observations of the brightest targets of our sample can be performed
by well-equipped amateur astronomers. The chances to detect circumbinary bodies does not depend only on the
precision of the individual timings but also on the number of participating institutions and devoted
amateurs and number of targets monitored.

The theoretical estimates show that the timing technique enables to detect circumbinary planets down to Jupiter
mass orbiting on a few-year orbits. The merit of an EB strongly depends on its brightness, depth, and width of
the minima (see Eqn.~\ref{precision_min}), less on the mass of the underlying EB (Eqn.~\ref{lite_amp}).

The observations within the project promise additional useful science such as: (i) the study of spot cycles in the RS~CVn-like
late-type binaries, detection of flares (see Pribulla et al., 2001), (ii) a more accurate characterization of recently-discovered
detached eclipsing binaries, (iii) detection of new low-mass EBs which is crucial to better define the empirical lower main
sequence, (iv) determination of absolute parameters of the components (in the case that spectroscopic orbits are available),
(v) detection of EBs with pulsating component(s), (vi) detection and characterization of multiple systems with two
systems of eclipses,
(vii) detection of new variable stars in the CCD fields covered, (viii) photometric detection of transits of
substellar components across the disks of the components of the eclipsing pair (see Doyle et al., 2011), (ix) detection of
invisible massive components causing precession of the EB orbit and changes of the minima depth (see Mayer
et al., 2004).

The LITE can always be regarded {\it only} as very good indication of a substellar body in the system. 
In nearby systems with a sufficiently close visual companion (e.g., CU~Cnc, GK~Boo) the LITE on a long-period
orbit could be possibly checked by the differential astrometry of the visual pair.


\acknowledgements
This work has been funded by the VEGA 2/0094/11. This study has made use of the SIMBAD database, operated at CDS,
Strasbourg, France and NASA's Astrophysics Data System Bibliographic Services. PSz and TSz would like to
thank the Hungarian OTKA Grant K76816. The work of the Bulgarian team (DK and DD) is partially supported by projects DO 02-362,
DO 02-85 and DDVU 02/40-2010 of the Bulgarian National Science Fund. JWL acknowledge the support of Korea Astronomy and
Space Science Institute (KASI) grant 2012-1-410-02. AF, GC, and AB gratefully acknowledge the Italian Ministero
dell'Istruzione, Universitae Ricerca (MIUR). PD would like to thank the Slovak Research and Development Agency grant
LPP-0024-09. GM and AN acknowledge Iuventus Plus grants IP2010 023070 and IP2011 031971. MZ would like to thank the
project MUNI/A/0968/2009. PO, ML and AH would like to thank the project FWF: P22950-N16 with title: "Stellar signatures
of mass expulsions in Radio and Optical wavelengths -- Importance for stellar planetary interactions".
MM acknowledge DFG for support in program MU2695/13-1. TP, MV, EK, and LH thank for the support to the project
APVV-0158-11. MAvE acknowledges support by DLR under the project 50 OW 0204. MHMM thanks to grant PEst-C/CTM/LA0025/2011 (FCT-Portugal).

\label{lastpage}
\end{document}